\documentclass[12pt]{article}

\usepackage{epsf}
\renewcommand{\thefootnote}{\fnsymbol{footnote}}

\begin{document}


\begin{titlepage}
\begin{flushright}
\begin{tabular}{l}
CERN--TH/99--377\\
DESY 99--191\\
hep-ph/9912319
\end{tabular}
\end{flushright}
\vskip1cm
\begin{center}
\boldmath
{\Large \bf An Analysis of $B_s$ Decays in the Left-Right-Symmetric

\vspace*{0.5truecm}

Model with Spontaneous CP Violation}
\unboldmath

\vspace{1.5cm}
{\sc Patricia~Ball}${}^{1,}$\footnote{E-mail: Patricia.Ball@cern.ch} and
{\sc Robert Fleischer}${}^{2,}$\footnote{E-mail: Robert.Fleischer@desy.de}
\\[0.5cm]
\vspace*{0.1cm} ${}^1${\it CERN-TH, CH--1211 Geneva 23, Switzerland
}\\[0.3cm]
\vspace*{0.1cm} ${}^2${\it Deutsches Elektronen-Synchrotron DESY, 
Notkestr.\ 85, D--22607 Hamburg, Germany}\\[0.8cm]


\vfill

{\large\bf Abstract\\[10pt]} \parbox[t]{\textwidth}{
Non-leptonic $B_s$ decays into CP eigenstates that are caused by 
$\bar b\to\bar cc\bar s$ quark-level transitions, such as 
$B_s\to D_s^+D^-_s$, $J/\psi\, \eta^{(')}$ or $J/\psi\, \phi$, provide 
a powerful tool to search for ``new physics'', as the CP-violating effects 
in these modes are tiny in the Standard Model. We explore these effects 
for a particular scenario of new physics, the left-right-symmetric model 
with spontaneous CP violation. In our analysis, we take into account all 
presently available experimental constraints on the parameters of this 
model, i.e.\ those implied by $K$- and $B$-decay observables; we find 
that CP asymmetries as large as ${\cal O}(40\%)$ may arise in the $B_s$ 
channels, whereas the left-right-symmetric model favours a small 
CP asymmetry in the ``gold-plated'' mode $B_d\to J/\psi\,K_{\rm S}$. 
Such a pattern would be in favour of $B$-physics experiments at hadron 
machines, where the $B_s$ modes are very accessible.}

  \vskip1.5cm

\end{center}

\end{titlepage}

\thispagestyle{empty}
\vbox{}
\newpage
 
\setcounter{page}{1}

\setcounter{footnote}{0}
\renewcommand{\thefootnote}{\arabic{footnote}}

\section{Introduction}\label{sec:intro}
A particularly interesting tool to search for indications of ``new physics''
is provided by $B_s$-meson decays into final CP eigenstates $|f\rangle$ 
that originate from $\bar b\to\bar cc\bar s$ quark-level transitions 
\cite{nirsil}--\cite{DFN}; important examples are given by 
$B_s\to D_s^+D^-_s$, $J/\psi\, \eta^{(')}$ or $J/\psi\, \phi$ decays. 
The interesting feature of these modes is that their decay amplitudes 
do not involve -- to a very good approximation -- a CP-violating weak 
phase in the Standard Model. Moreover, the weak $B^0_s$--$\overline{B^0_s}$ 
mixing phase, which governs ``mixing-induced'' CP violation, is negligibly 
small in the Standard Model. Consequently, these $B_s$ decays exhibit tiny 
CP-violating effects within the Kobayashi--Maskawa picture of CP violation, 
thereby representing a sensitive probe for CP-violating contributions 
from physics beyond the Standard Model. 

We analyse these effects for a particular scenario of new physics, the 
symmetrical $SU(2)_{\rm L}\times SU(2)_{\rm R}\times U(1)$ model with 
spontaneous CP violation (SB--LR) \cite{LR-refs,JMF}. In a recent paper 
\cite{my}, the SB--LR model has been investigated in the light of current 
experimental constraints from $K$- and $B$-decay observables. In a
large region of parameter space, the model mainly affects neutral-meson 
mixing, but does not introduce sizeable ``direct'' CP violation. The 
sensitive observables constraining the model are thus the meson mass 
differences $\Delta M_K$, $\Delta M_{B_d}$, $\Delta M_{B_s}$, the ``indirect'' 
CP-violating parameter $\epsilon_K$ of the neutral kaon system, and 
the mixing-induced CP asymmetry 
${\cal A}^{\rm mix}_{\rm CP}(B_d\to J/\psi\, K_{\rm S})$.
In particular, it was found that, for a set of fixed CKM parameters and 
quark masses, the model predicts a small value for 
$|{\cal A}^{\rm mix}_{\rm CP}(B_d\to J/\psi\, K_{\rm S})|$ below $10\%$,
which is in agreement at the 2$\sigma$ level with the CDF measurement
$0.79^{+0.41}_{-0.44}$ \cite{CDF}, but at variance with the Standard-Model 
expectation $0.73\pm0.21$ \cite{AL}. 

The new point we want to make in this letter is that the SB--LR model 
predicts  values also for the mixing-induced CP asymmetries of the $B_s$
decays considered here, for example $B_s\to J/\psi\, \phi$, that 
largely deviate from the Standard-Model expectation of very small 
CP-violating effects. We show that mixing-induced CP asymmetries as large 
as ${\cal O}(40\%)$ may arise in these channels, whereas direct CP violation 
stays very small. We thus face the interesting possibility that, with all 
current experimental constraints being met, new physics may just lurk around 
the corner, and may be revealed by the pattern of CP violation exhibited by 
$B_d\to J/\psi\, K_{\rm S}$ and $B_s\to D_s^+D^-_s$, $J/\psi\, \eta^{(')}$ 
or $J/\psi\, \phi$. This scenario, with small CP-violating effects in the
former decay and large effects in the latter ones, would be in favour 
of $B$-physics experiments at hadron machines, where the $B_s$ modes are 
very accessible, in contrast to the situation at asymmetric $e^+$--$e^-$
$B$-factories operating on the $\Upsilon(4S)$ resonance.

The outline of this paper is as follows: in Section~\ref{sec:AO}, we have
a brief look at the structure of the Standard-Model decay amplitudes of 
the $B_s$-meson decays considered here, and introduce the corresponding 
CP-violating observables. The basic features of the left-right-symmetric 
model with spontaneous CP violation are discussed in 
Section~\ref{sec:LR-Model}, and the numerical analysis is presented in 
Section~\ref{sec:analysis}. Finally, in 
Section~\ref{sec:concl} we summarize our conclusions. 
\section{Decay Amplitudes and CP-Violating Observables}\label{sec:AO}
Before we introduce the CP-violating observables, let us have a brief 
look at the structure of the Standard-Model transition amplitudes of 
$B_s$ decays of the kind $B_s\to D_s^+D^-_s$, $J/\psi\, \eta^{(')}$ 
or $J/\psi\, \phi$. The new-physics contributions to the decay amplitudes 
of these channels arising within the left-right-symmetric model with 
spontaneous CP violation will be discussed in Section~\ref{sec:LR-Model}.
\subsection{The Standard-Model Decay Amplitudes}\label{sec:SM-ampls}

Within the Standard Model, the amplitudes of $B^0_s$-meson decays caused
by $\bar b\to\bar cc\bar s$ quark-level transitions can be expressed 
generically as follows \cite{rev}:
\begin{equation}\label{gen-ampl}
A(B_s^0\to f)=\lambda_c^{(s)}\left(A_{\rm cc}^{c}+
A_{\rm pen}^{c}\right)+\lambda_u^{(s)}A_{\rm pen}^{u}
+\lambda_t^{(s)}A_{\rm pen}^{t}\,,
\end{equation}
where $f\in\{D_s^+D_s^-, J/\psi\, \eta^{(')}, \ldots \}$ is a final-state
configuration with $\bar cc\bar s s$ valence-quark content, $A_{\rm cc}^{c}$ 
denotes current--current contributions, i.e.\ ``tree'' processes, and the 
amplitudes $A_{\rm pen}^{q}$ describe the contributions from penguin 
topologies with internal $q$ quarks ($q\in\{u,c,t\})$. These penguin 
amplitudes take into account both QCD and electroweak penguin 
contributions \cite{rev}. The $\lambda_q^{(s)}\equiv V_{qs}V_{qb}^\ast$
are the usual CKM factors. Making use of the unitarity of the CKM matrix
and applying the Wolfenstein parametrization \cite{wolf}, generalized to
include non-leading terms in $\lambda$ \cite{blo}, we obtain~\cite{RF-psiK}
\begin{equation}\label{Bd-ampl2}
A(B_s^0\to f)=\left(1-\frac{\lambda^2}{2}\right){\cal A}
\left[1+\left(\frac{\lambda^2}{1-\lambda^2}\right)a e^{i\theta}e^{i\gamma}
\right],
\end{equation}
where
\begin{equation}\label{Aap-def}
{\cal A}\equiv\lambda^2A\left(A_{\rm cc}^{c}+A_{\rm pen}^{ct}\right),
\end{equation}
with $A_{\rm pen}^{ct}\equiv A_{\rm pen}^{c}-A_{\rm pen}^{t}$, and
\begin{equation}\label{ap-def}
ae^{i\theta}\equiv R_b\left(\frac{A_{\rm pen}^{ut}}{A_{\rm cc}^{c}+
A_{\rm pen}^{ct}}\right).
\end{equation}
The quantity $A_{\rm pen}^{ut}$ is defined in analogy to $A_{\rm pen}^{ct}$,
and the relevant CKM factors are given by
\begin{equation}\label{CKM-exp}
\lambda\equiv|V_{us}|=0.22\,,\quad A\equiv\frac{1}{\lambda^2}
\left|V_{cb}\right|=0.81\pm0.06\,,\quad R_b\equiv
\left(1-\frac{\lambda^2}{2}\right)\frac{1}{\lambda}
\left|\frac{V_{ub}}{V_{cb}}\right|=0.41\pm0.07\,.
\end{equation}
In Eq.\ (\ref{Bd-ampl2}), the CP-violating weak phase $\gamma$ is the 
usual angle of the unitarity triangle of the CKM matrix, whereas $\theta$ 
denotes a CP-conserving strong phase. 
\boldmath
\subsection{The CP-Violating Observables}\label{Obs}
\unboldmath
For a $B_s$ decay into a final CP eigenstate $|f\rangle$, such as 
$B_s\to D_s^+D^-_s$ or $J/\psi\, \eta^{(')}$, $B^0_s$--$\overline{B^0_s}$ 
oscillations lead to the following time-dependent CP asymmetry:
\begin{eqnarray}
a_{\rm CP}(t) & \equiv & \frac{\Gamma(B^0_s(t)\to f)-
\Gamma(\overline{B^0_s}(t)\to f)}{\Gamma(B^0_s(t)\to f)+
\Gamma(\overline{B^0_s}(t)\to f)}\nonumber\\
 &=& 2\,e^{-\Gamma_s t}\left[\frac{{\cal A}_{\rm CP}^{\rm dir}(B_s\to f)
\cos(\Delta M_s t)+{\cal A}_{\rm CP}^{\rm mix}(B_s\to f)\sin(\Delta M_s t)}{
e^{-\Gamma_{\rm H}^{(s)}t}+e^{-\Gamma_{\rm L}^{(s)}t}+
{\cal A}_{\rm \Delta\Gamma}(B_s\to f)\left(e^{-\Gamma_{\rm H}^{(s)}t}-
e^{-\Gamma_{\rm L}^{(s)}t}\right)} \right],\label{ee6}
\end{eqnarray}
where $\Delta M_s\equiv M_{\rm H}^{(s)}-M_{\rm L}^{(s)}$ denotes the
mass difference between the $B_s$ mass eigenstates $B_s^{\rm H}$ (``heavy'')
and $B_s^{\rm L}$ (``light''), the $\Gamma_{\rm H,L}^{(s)}$ are the 
corresponding decay widths, and $\Gamma_s$ is defined as 
$\Gamma_s\equiv\left[\Gamma_{\rm H}^{(s)}+\Gamma_{\rm L}^{(s)}\right]/2$. 
In Eq.\ (\ref{ee6}), we have separated the ``direct'' from the 
``mixing-induced'' CP-violating contributions, which are described by
${\cal A}_{\rm CP}^{\rm dir}(B_s\to f)$ and ${\cal A}_{\rm CP}^{\rm 
mix}(B_s\to f)$, respectively \cite{rev}. In contrast to the $B_d$ system, 
the width difference 
\begin{equation}\label{DG-def}
\Delta\Gamma_q\equiv\Gamma_{\rm H}^{(q)}-\Gamma_{\rm L}^{(q)}
\end{equation}
may be sizeable in the $B_s$ system \cite{DGamma-cal}, thereby providing 
the observable ${\cal A}_{\rm \Delta\Gamma}(B_s\to f)$. This quantity is 
not independent from ${\cal A}^{\mbox{{\scriptsize dir}}}_{\mbox{{\scriptsize 
CP}}}(B_s\to f)$ and ${\cal A}^{\mbox{{\scriptsize 
mix}}}_{\mbox{{\scriptsize CP}}}(B_s\to f)$, but satisfies the following 
relation:
\begin{equation}\label{Obs-rel}
\Bigl[{\cal A}_{\rm CP}^{\rm dir}(B_s\to f)\Bigr]^2+
\Bigl[{\cal A}_{\rm CP}^{\rm mix}(B_s\to f)\Bigr]^2+
\Bigl[{\cal A}_{\Delta\Gamma}(B_s\to f)\Bigr]^2=1.
\end{equation}
Interestingly, the observable ${\cal A}_{\rm \Delta\Gamma}(B_s\to f)$
can be extracted from CP-violating effects in ``untagged'' $B_s$ 
rates \cite{dun,FD}:
\begin{equation}\label{untagged}
\Gamma[f(t)]\equiv\Gamma(B^0_s(t)\to f)+\Gamma(\overline{B^0_s}(t)
\to f)\propto R_{\rm H}(B_s\to f)\,e^{-\Gamma_{\rm H}^{(s)}t}+
R_{\rm L}(B_s\to f)\,e^{-\Gamma_{\rm L}^{(s)}t},
\end{equation}
where
\begin{equation}\label{RHL-def}
R_{\rm H}(B_s\to f)=\frac{1}{2}\left[1+{\cal A}_{\rm \Delta\Gamma}(B_s\to f)
\right],\quad R_{\rm L}(B_s\to f)=\frac{1}{2}\left[1-
{\cal A}_{\rm \Delta\Gamma}(B_s\to f)\right],
\end{equation}
and hence
\begin{equation}\label{ADG}
{\cal A}_{\Delta\Gamma}(B_s\to f)=\frac{R_{\rm H}(B_s\to f)-
R_{\rm L}(B_s\to f)}{R_{\rm H}(B_s\to f)+R_{\rm L}(B_s\to f)}\,.
\end{equation}
Studies of such untagged rates, where there are no rapid oscillatory 
$\Delta M_st$ terms present, are more promising than tagged rates 
in terms of efficiency, acceptance and purity.

Looking at (\ref{Bd-ampl2}), we observe that the weak phase factor 
$e^{i\gamma}$, which is associated with the ``penguin parameter'' 
$a e^{i\theta}$, is strongly Cabibbo-suppressed by $\lambda^2$. 
Consequently, there is no CP-violating weak phase present in this decay 
amplitude to an excellent approximation. In this very important special 
case, we obtain (for details, see \cite{DFN,rev}): 
\begin{equation}\label{BPP-obs}
{\cal A}^{\mbox{{\scriptsize dir}}}_{\mbox{{\scriptsize CP}}}(B_s\to f)=
0\,,\quad
{\cal A}^{\mbox{{\scriptsize mix}}}_{\mbox{{\scriptsize CP}}}(B_s\to f)=
\sin\phi_s\,,\quad
{\cal A}_{\rm \Delta\Gamma}(B_s\to f)=-\cos\phi_s\,,
\end{equation}
where $\phi_s$ denotes a phase-convention-independent combination of the
CP-violating weak $B^0_s$--$\overline{B^0_s}$ mixing and 
$\bar b\to\bar cc\bar s$ decay phases. In general, we have
\begin{equation}
\phi_s=\phi_s^{\rm SM}+\phi_s^{\rm NP}, 
\end{equation}
where the Standard-Model phase 
\begin{equation}
\phi_s^{\rm SM}=2\,\mbox{arg}(-V_{ts}^\ast V_{tb}^{\vphantom{*}})+
2\,\mbox{arg}(V_{cs}^{\vphantom{*}}V_{cb}^\ast)=-\,2\lambda^2\eta\approx-0.03
\end{equation}
is negligibly small, and $\phi_s^{\rm NP}$ is due to new physics.
Within the Standard Model, the CP-violating effects in the $B_s$ decays 
considered here are thus very small. However, $\phi_s^{\rm NP}$ may be 
sizeable in our scenario for new physics, as we will see in 
Section~\ref{sec:analysis}, thereby leading to significant mixing-induced
CP violation. A similar feature arises also in some other scenarios for 
physics beyond the Standard Model, for example in models allowing mixing 
to a new isosinglet down quark, as in E$_6$ \cite{silver}. Unfortunately,
the new-physics effects reduce the magnitude of the 
$B_s^0$--$\overline{B_s^0}$ width difference as follows \cite{grossman}:
\begin{equation}\label{DGamNP}
\Delta\Gamma_s=\Delta\Gamma_s^{\rm SM}\cos\phi_s,
\end{equation}
where $\Delta\Gamma_s^{\rm SM}={\cal O}(-15\%)$ is the Standard-Model 
width difference \cite{DGamma-cal}. Note that (\ref{DG-def}) implies
a negative Standard-Model width difference. However, the sign of 
$\Delta\Gamma_s$ may change in the presence of new physics, as can be 
seen in (\ref{DGamNP}).

The situation in the decay $B_s\to J/\psi\, \phi$, which is very 
promising for $B$-physics experiments at hadron machines because 
of its favourable experimental signature, is a bit more involved 
than in the case of the pseudoscalar--pseudoscalar modes $B_s\to D_s^+D^-_s$ 
and $J/\psi\, \eta^{(')}$, since the final state is an admixture of different 
CP eigenstates. In the case of decays into two vector mesons, such as
$B_s\to J/\psi\, \phi$, it is convenient to introduce linear polarization 
amplitudes $A_0(t)$, $A_\parallel(t)$ and $A_\perp(t)$ \cite{pol}. Whereas 
$A_\perp(t)$ describes a CP-odd final-state configuration, both $A_0(t)$ 
and $A_\parallel(t)$ correspond to CP-even final-state configurations, i.e.\
to the CP eigenvalues $-1$ and $+1$, respectively. In order to disentangle 
them, one has to study angular distributions of the decay products 
of the decay chain $B_s\to J/\psi[\to l^+l^-]\, \phi[\to K^+K^-]$, which can 
be found in \cite{ddf1}.\footnote{For a detailed discussion of new-physics 
effects in the corresponding observables, see \cite{DFN}.} Let us here 
just give the following time-dependent CP asymmetry, under the same 
assumption as was made in (\ref{BPP-obs}), i.e.\ that there is no CP-violating 
weak phase present in the $B_s\to J/\psi\, \phi$ decay amplitude:
\begin{equation}\label{CP-asym}
a_{\rm CP}(B_s(t)\to J/\psi\,\phi)\equiv\frac{\Gamma(t)-
\overline{\Gamma}(t)}{\Gamma(t)+\overline{\Gamma}(t)}
=\left[\frac{1-D}{F_+(t)+D F_-(t)}\right]
\sin(\Delta M_s t)\,\sin\phi_s,
\end{equation}
where $\Gamma(t)$ and $\overline{\Gamma}(t)$
denote the time-dependent rates for decays of initially, i.e.\ at $t=0$, 
present $B^0_s$- and $\overline{B^0_s}$-mesons into $J/\psi\,\phi$ final 
states, respectively. The remaining quantities are defined as 
\begin{equation}
D\equiv\frac{|A_{\perp}(0)|^2}{|A_0(0)|^2 + |A_{\|}(0)|^2}\,,
\end{equation}
and
\begin{equation}
F_{\pm}(t)\equiv\frac{1}{2}\left[\left(1\pm\cos\phi_s\right)
e^{+\Delta\Gamma_s t/2}+\left(1\mp\cos\phi_s\right)
e^{-\Delta\Gamma_s t/2}\right].
\end{equation}
Note that we have $F_+(t)=F_-(t)=1$ for a negligible width difference 
$\Delta\Gamma_s$. Obviously, the advantage of the ``integrated'' observable 
(\ref{CP-asym}) is that it can be measured {\it without} performing an
angular analysis. The disadvantage is of course that -- in 
contrast to (\ref{BPP-obs}) -- it  also depends on the hadronic quantity 
$D$, which precludes a theoretically clean extraction of $\phi_s$ from 
(\ref{CP-asym}). However, this feature does not limit the power of this 
CP asymmetry to search for indications of new physics, which would be 
provided by a measured sizeable value of (\ref{CP-asym}). Model calculations 
of $D$, making use of the factorization hypothesis, typically give 
$D=0.1\ldots0.5$ \cite{ddf1}, which is also in agreement with a recent 
analysis of the $B_s\to J/\psi\, \phi$ polarization amplitudes performed
by the CDF collaboration \cite{CDF-schmidt}. A recent calculation of
the relevant hadronic form factors from QCD sum rules on the
light-cone \cite{BB} yields $D=0.33$ in the factorization approximation. 
Consequently, the CP-odd 
contributions proportional to $|A_{\perp}(0)|^2$ may have a significant 
impact on (\ref{CP-asym}). In order to extract $\phi_s$ from CP-violating 
effects in the decay $B_s\to J/\psi\, \phi$ in a theoretically clean way, 
an angular analysis has to be performed, as is discussed in detail 
in \cite{DFN}.  
\section{The Left-Right-Symmetric Model with Spontaneous CP 
Violation}\label{sec:LR-Model}
Before discussing its predictions for CP-violating phenomena, let us
explain very shortly the essential features of the SB--LR model. It
is based on the gauge group $SU(2)_{\rm R}\times SU(2)_{\rm L} \times
U(1)$, which cascades down to the unbroken electromagnetic subgroup
$U(1)_{\rm em}$ through the following simple symmetry-breaking
pattern:
$$
\underbrace{SU(2)_{\rm R}\times SU(2)_{\rm L}}_{
\underbrace{\displaystyle SU(2)_{\rm L}\hspace*{0.6cm}
  \times U(1)}_{\displaystyle U(1)_{\rm em}}\hspace*{-2.4cm}}\times U(1)
$$
The scalar sector is highly model-dependent; for the
generation of quark masses, there has to be
 at least one scalar bidoublet $\Phi$, i.e.\ a
doublet under both $SU(2)$, which, by spontaneous breakdown of
$SU(2)_{\rm R}\times SU(2)_{\rm L}$, acquires the VEV
\begin{equation}\label{eq:VEV}
\langle \Phi \rangle = \frac{1}{\sqrt{2}}\left( \begin{array}{cc} v &
    0\\ 0 & w\end{array}\right).
\end{equation}
In general, both $v$ and $w$ are complex, which is the (only) source of
CP violation in the model. The particle content of $\Phi$ corresponds to 
four particles, one analogue of the Standard-Model Higgs, two 
flavour-changing neutral Higgs bosons, and one flavour-changing charged 
Higgs. The masses of these new Higgs particles can be assumed to be 
degenerate to good accuracy, and were found to lie in the range
$10.2\,{\rm TeV}< M_H < 14.6\,{\rm TeV}$ \cite{my}.

LR symmetry implies that the left-handed quark sector of the Standard Model 
gets complemented by a right-handed one, with quark mixing matrices 
$V_{\rm L}$ and $V_{\rm R}$, respectively, and $|V_{\rm L}|=|V_{\rm
  R}|$. In the standard Maiani convention, 
$V_{\rm L}$ contains one, $V_{\rm R}$ five complex phases, which depend on 
the three generalized Cabibbo-type angles (``CKM angles''), the quark masses, 
and the VEV (\ref{eq:VEV}). The presence of such a large number of weak 
phases, calculable in terms of only one non-Standard-Model variable, is what 
makes the investigation of CP-violating phenomena in the SB--LR model that 
interesting. The left- and right-handed charged gauge bosons $W_{\rm L}$ and 
$W_{\rm R}$ mix with each other;  the mass of the predominantly right-handed 
mass eigenstate $W_2$, $M_2$, is found to lie in the range 
$2.75\,{\rm TeV} < M_2 < 13\,{\rm TeV}$ \cite{my}. The mixing angle 
$\zeta$, defined as
\begin{equation}
\zeta = \frac{2 |vw|}{|v|^2 + |w|^2}\left( \frac{M_1}{M_2}\right)^2,
\end{equation}
is rather small: as the 
ratio $|v|/|w|$ is smaller than 1,\footnote{Which can always be
achieved by a redefinition of the Higgs bidoublet $\Phi\to \sigma_2
\Phi^* \sigma_2$.} one has $\zeta < (M_1/M_2)^2 = 8.5\times 10^{-4}$.
There are, however, arguments, according to which a small ratio
$|v|/|w|\sim {\cal O}(m_b/m_t)$ would naturally explain the observed
smallness of the CKM angles \cite{natural}; in this case, $\zeta < 0.3
\times 10^{-4}$. An experimental bound on $\zeta$ 
can in principle be obtained from the
upper bound on the electromagnetic dipole moment of the neutron, which
is due to L--R mixing; existing theoretical calculations are,
however, very sensitive to the precise values of only poorly known
nucleon matrix elements;\footnote{In addition, the Higgs contributions
  to the dipole moment are usually not included.}
 the present status of an
experimental bound on $\zeta$ is thus not quite clear, although large values
of $\zeta \sim {\cal O}(10^{-4})$ appear to be disfavoured (see also the
discussion in \cite{my}).
The fact that the new boson masses are in the TeV range implies 
that the SB--LR model has no perceptible impact on Standard-Model tree-level 
amplitudes; rather, it manifests itself in
\begin{itemize}
\item $W_{\rm L}$--$W_{\rm R}$ mixing in top-dominated penguin diagrams, 
enhanced by large quark mass terms from spin-flips, $\zeta\to \zeta\,
  m_t/m_b$ (similar for penguins with charged Higgs particles),
\item Standard-Model amplitudes that are forbidden or heavily suppressed 
  (electric
  dipole moment of the neutron, for instance),
\item mixing of neutral $K$- and $B$-mesons, where the suppression
  factor of $(M_1/M_2)^2$ is partially compensated by large
  Wilson coefficients or hadronic matrix elements (chiral enhancement
  in $K$ mixing), and to which the flavour-changing Higgs bosons contribute 
  at tree level. 
\end{itemize}
We thus expect the SB--LR model to change the $B_s^0\to f$ amplitude 
defined in (\ref{Bd-ampl2}) in the following way:
\begin{equation}
{\cal A} \to {\cal A}^{\rm LR} =  \lambda^2 A \left(A^c_{cc} + A^{c,{\rm
    LR}}_{cc} + A^{ct}_{\rm pen} + A^{ct, {\rm LR}}_{\rm
    pen}\right),
\end{equation}
with
\begin{equation}
A^{c,{\rm LR}}_{cc}  \sim  {\cal O}\left(
    \left[\frac{M_1}{M_2}\right]^2 A_{cc}^c\right),\quad
A^{ct,{\rm LR}}_{\rm pen}  \sim  {\cal O}\left( \zeta 
    \left[\frac{m_c}{m_b}\, A^c_{\rm pen} - \frac{m_t}{m_b}\, 
A^t_{\rm pen}\right]\right).
\end{equation}
Numerically, $(M_1/M_2)^2 < 10^{-3}$ and $\zeta\, m_t/m_b < 0.05$ for
maximum $|v|/|w|=1$, and $\zeta\, m_t/m_b < 6\times 10^{-3}$ for the more
likely case of $|v|/|w|\sim {\cal O}(m_b/m_t)$. In any case, it is clear
that the specific LR contributions to the decay amplitude, although
they carry new weak phases, are heavily suppressed by powers of $M_2$
(and $M_H$ for the corresponding charged Higgs penguins).
Consequently, the new contributions of the SB--LR model to the amplitudes 
of the $B$ decays considered in this letter are small and
do not yield sizeable direct CP violation. The assumption used to 
calculate the CP-violating observables (\ref{BPP-obs}) and (\ref{CP-asym}) 
is therefore also satisfied in the SB--LR model, so that we may use these
expressions in the numerical analysis given in the following section. 
\section{Numerical Analysis}\label{sec:analysis}
In the SB--LR model, the Standard-Model mixing matrix gets modified by 
$W_{\rm R}$ boxes and tree-level flavour-changing neutral Higgs exchange as
\begin{equation}
M_{12}^{B_q} = M_{12}^{\rm SM} + M_{12}^{\rm LR} \equiv M_{12}^{\rm SM} ( 1 +
\kappa e^{i\sigma_q})
\end{equation}
with
\begin{equation}
\kappa \equiv \left|\frac{M_{12}^{\rm LR}}{M_{12}^{\rm SM}}\right|, \quad
 \sigma_q \equiv \arg \,\frac{M_{12}^{\rm LR}}{M_{12}^{\rm SM}} =
 \arg\left( -\frac{V_{tb}^R V_{tq}^{R*}}{V_{tb}^L V_{tq}^{L*}}\right),
\end{equation}
such that the relevant observable phase $\phi_q$ becomes
\begin{equation}
  \phi_q = \phi^{\rm SM}_q + \arg \left( 1 + \kappa
  e^{i\sigma_q}\right),
\end{equation} 
where $q\in\{d,s\}$. The $B_d$ counterpart $\phi_d$ to the phase $\phi_s$ 
can be determined in a theoretically clean way through mixing-induced CP 
violation in the ``gold-plated'' mode $B_d\to J/\psi\,K_{\rm S}$ \cite{bisa}:
\begin{equation}
{\cal A}^{\rm mix}_{\rm CP}(B_d\to J/\psi\, K_{\rm S})=-\sin\phi_d\,.
\end{equation}
To good accuracy, $\kappa$ is independent on the flavour of the
spectator quark $q$, and is given as \cite{my}
\begin{equation}
\kappa = (1.2\pm 0.2) \left[ \left( \frac{7\,{\rm TeV}}{M_H}\right)^2
  + 1.7 \left(\frac{1.6\,{\rm TeV}}{M_2}\right)^2 \left\{ 0.051 -
  0.013 \ln \left(\frac{1.6\,{\rm TeV}}{M_2}\right)^2\right\}\right].
\end{equation}
A crucial consequence of the spontaneous breakdown of the CP symmetry is
that the phases of the two quark mixing matrices, and hence also
$\sigma_q$, can be calculated in
terms of the quark masses, the three CKM angles and the VEV value of
$\Phi$, Eq.~(\ref{eq:VEV}); they do not depend on $M_2$ or $M_H$. 
For the technicalities, we refer to Ref.~\cite{my}; here we only state
that the dependence on $\langle \Phi\rangle$ can be lumped into 
 a single
variable, $\beta$, which is defined as
\cite{JMF} 
\begin{equation}
\beta = \arctan \,\frac{2 |wv| \sin [\arg(vw)]}{|v|^2-|w|^2}\,.
\end{equation}
{}From the requirement that diagonalization of the quark
mass matrices be possible, $\beta$ is bounded as follows \cite{JMF}:
$$
\tan\,\frac{\beta}{2} \leq \frac{m_b}{m_t}.
$$
The fact that quark mass signs are observable in the SB--LR model
entails a 64-fold discrete ambiguity of the complex phases of $V_{\rm L}$ and
$V_{\rm R}$. The dependence of the phases on the input parameters 
can be obtained in analytical form in a
linear expansion in $\beta$, the so-called small-phase approximation
\cite{LR-refs}, which is appropriate for studying the $K$ system. For
$B$ decays, however, the approximation breaks down; the full
functional dependence of $\sigma_q$ on $\beta$  
can only be obtained numerically and has been calculated in \cite{my}.

Experimental observables sensitive to the new-physics contributions to
the off-diagonal element $M_{12}$ of the mixing matrix are, apart from 
${\cal A}^{\rm mix}_{\rm CP}(B_d\to J/\psi\, K_{\rm S})$, in particular 
the meson mass differences $\Delta M_K$, $\Delta M_{B_d}$ and 
$\Delta M_{B_s}$, which are given by $\Delta M = 2 |M_{12}|$. Other relevant 
observables are $\epsilon_K$, which is sensitive to $\arg M_{12}^K$, the 
observable Re$(\epsilon'_K/\epsilon_K)$ measuring direct CP violation in the
neutral kaon system,\footnote{Because of the large hadronic
uncertainties affecting Re$(\epsilon'_K/\epsilon_K)$, it has only been used
that a positive value of this observable is implied by the most recent
experimental data \cite{epsprime}.} and the upper bound on the 
electromagnetic dipole moment of the neutron. Without going into details 
about the respective strength and viability of these constraints, we 
just quote the results of the comprehensive analysis of Ref.~\cite{my}: 
using the following set of Standard-Model input parameters,
\begin{equation}\label{eq:input}
\renewcommand{\arraystretch}{1.4}
\begin{array}[b]{lcllcllcl}
\overline{m}_t(\overline{m}_t) & = & 170\,{\rm GeV}, &
\overline{m}_b(\overline{m}_b) & = & 4.25\,{\rm GeV},\\
\overline{m}_c(\overline{m}_c) & = & 1.33\,{\rm GeV}, & 
\overline{m}_s(2\,{\rm GeV}) & =& 110\,{\rm MeV},\\
m_s/m_d & = & 20.1, & m_u/m_d & = & 0.56,\\
\multicolumn{6}{l}{|V_{us}|  =  0.2219,\quad |V_{ub}|  =  0.004,\quad
|V_{cb}| = 0.04,}
\end{array}
\renewcommand{\arraystretch}{1}
\end{equation}
and neglecting their uncertainties, the 64-fold phase ambiguity gets 
completely resolved by requiring both $\sin[\arg M_{12}^K]$ and $\sin\phi_d$
to be positive, as implied by the measured values 
of $\epsilon_K$ and ${\cal A}^{\rm mix}_{\rm CP}(B_d\to
J/\psi\, K_{\rm S})$. Choosing the SB--LR model parameters $M_2$,
$M_H$ and $\beta$ such as to reproduce the experimental value of
$\epsilon_K$, one finds that the predicted value
$-0.1 < {\cal A}^{\rm mix}_{\rm CP}(B_d\to J/\psi\, K_{\rm S})<0.1$ is
rather small (see Fig.~9 in Ref.~\cite{my}) but still in agreement at the 
2$\sigma$ level with the CDF measurement $-0.79^{+0.44}_{-0.41}$ \cite{CDF}.
The parameter $\kappa$ is found to lie in the interval [0.29,0.74]; 
$\sigma_d$ assumes values in [0.12,1.81] rad and $\sigma_s$ in 
[3.23,4.58] rad. 

\begin{figure}
$$\epsfxsize=0.5\textwidth\epsffile{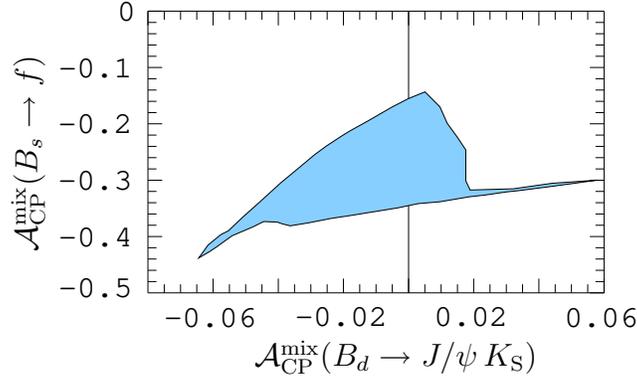}$$
\caption[]{The allowed region in the space of ${\cal A}^{\rm mix}_{\rm CP}
(B_d\to J/\psi\, K_{\rm S})=-\sin\phi_d$ and 
${\cal A}^{\rm mix}_{\rm CP}(B_s\to f)=\sin\phi_s$, with 
$f=D_s^+D^-_s$, $J/\psi\, \eta^{(')}$, in the SB--LR model.}\label{fig:corr1}
\end{figure}

\begin{figure}
$$\epsfxsize=0.5\textwidth\epsffile{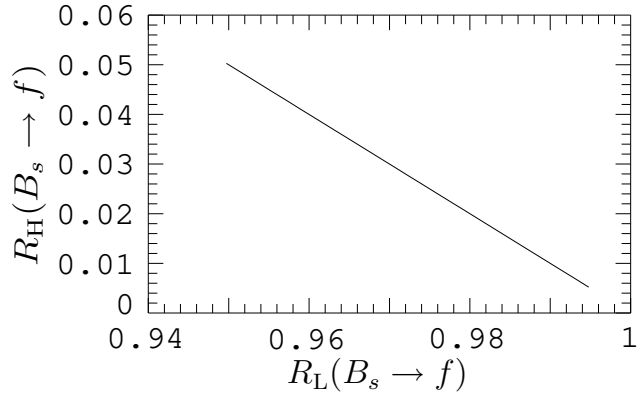}$$
\caption[]{The correlation between the observables $R_{\rm L}(B_s\to f)$
and $R_{\rm H}(B_s\to f)$ of the untagged $B_s\to f$ rates, with 
$f=D_s^+D^-_s$, $J/\psi\, \eta^{(')}$, in the 
SB--LR model.}\label{fig:corr-untagged}
\end{figure}

We are now in a position to calculate the correlation between the 
mixing-induced CP asymmetries in the decays $B_d\to J/\psi\, K_{\rm S}$ and 
$B_s\to D_s^+ D_s^-$, $J/\psi\,\eta^{(')}$ within the SB--LR model. 
The result is given in Fig.~\ref{fig:corr1}, which nicely shows 
that large CP-violating asymmetries in $B_s\to D_s^+D^-_s$, 
$J/\psi\, \eta^{(')}$ are possible in the SB--LR model, whereas 
mixing-induced CP violation in $B_d\to J/\psi\, K_{\rm S}$ is predicted 
to be small. As we have already noted in Section~\ref{sec:LR-Model}, the 
corresponding direct CP asymmetries remain very small, since the new 
contributions of the SB--LR model to the decay amplitudes are strongly 
suppressed. 

\begin{figure}
$$\epsfxsize=0.5\textwidth\epsffile{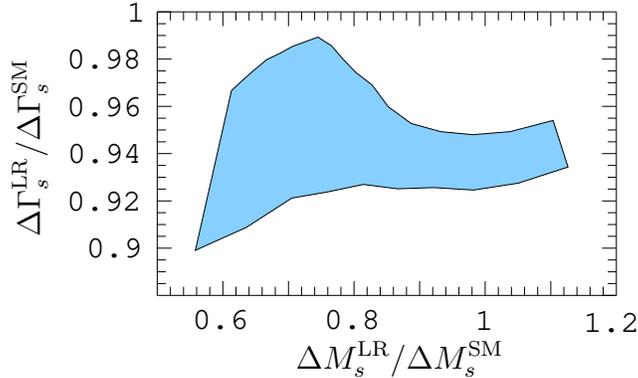}$$
\caption[]{The correlation between the $B^0_s$--$\overline{B^0_s}$ mass 
and width differences, normalized to their Standard-Model values, in the 
SB--LR model.}\label{fig:corr2}
\end{figure}

In Fig.~\ref{fig:corr-untagged}, we show the correlation between the 
observables of the untagged $B_s\to f$ rates ($f=D_s^+D^-_s$, $J/\psi\, 
\eta^{(')}$), which were introduced in (\ref{untagged}), and are given by
\begin{equation}
R_{\rm L}(B_s\to f)=\frac{1}{2}\left(1+\cos\phi_s\right),\quad
R_{\rm H}(B_s\to f)=\frac{1}{2}\left(1-\cos\phi_s\right).
\end{equation}
As can be seen there, in the case of the SB--LR model,
the component entering with $e^{-\Gamma_{\rm H}^{(s)}t}$ is at most 
$10\%$ of that associated with $e^{-\Gamma_{\rm L}^{(s)}t}$.
In order to extract $R_{\rm L}(B_s\to f)$ and $R_{\rm H}(B_s\to f)$, 
a sizeable value of $\Delta\Gamma_s$ is required, as we already noted 
in Section~\ref{sec:AO}. In Fig.~\ref{fig:corr2}, we show the correlation 
between the $B^0_s$--$\overline{B^0_s}$ mass and width differences in the 
SB--LR model. The reduction of $\Delta\Gamma_s$ through new-physics effects, 
which is described by (\ref{DGamNP}), is fortunately not very effective 
in this case, whereas the mass difference $\Delta M_s$ may be reduced
significantly. Although, at first glance, values of $\Delta M_s$ as small 
as $0.55\Delta M_s^{\rm SM}$ may seem to be at variance with the 
experimental bound of $\Delta M_s > 14.3\,{\rm ps}^{-1}$ at 95\% C.L.\ 
\cite{blaylock}, this is actually 
not the case: with the hadronic parameters from \cite{laurent} and 
$|V_{ts}| = 0.04$ with the generalized Cabibbo-angles fixed from 
(\ref{eq:input}), one has the theoretical prediction (see \cite{my},
e.g., for the full formula)
$$
\Delta M_s^{\rm SM} = (14.5\pm 6.3) {\rm ps}^{-1}.
$$
Combining this with the experimental bound, one has 
$$
\frac{\Delta M_s^{\rm LR}}{\Delta M_s^{\rm SM}} > \frac{14.3}{14.5 + 2
  \times 6.3} = 0.53.
$$
A pattern of $B_s$ mass and decay width differences like that emerging in the 
SB--LR model would be in favour of experimental studies of the $B_s$ decays
at hadron machines, where small values of $\Delta M_s$ and large values of
$\Delta\Gamma_s$ would be desirable. Because of the small ratio of
$R_{\rm H}(B_s\to f)/R_{\rm L}(B_s\to f)<0.1$ of the ``untagged'' $B_s\to f$
observables, ``tagged'' studies, allowing us to extract the mixing-induced
CP asymmetries ${\cal A}^{\rm mix}_{\rm CP}(B_s\to f)$, appear more 
promising to search for indications of the SB--LR model. However, in other 
scenarios for new physics, the situation may be different. 

\begin{figure}
$$\epsfxsize=0.5\textwidth\epsffile{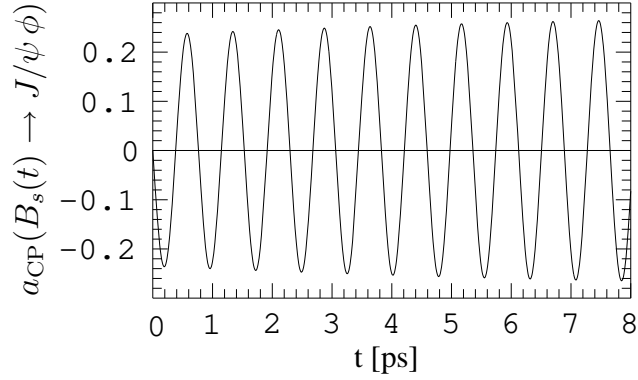}$$
\caption[]{The time-dependent CP-asymmetry $a_{\rm CP}(B_s(t)\to J/\psi\,
  \phi)$ introduced in (\protect{\ref{CP-asym}}) for fixed values of
$D=0.3$, $\sin\phi_s = -0.38$, $\Delta \Gamma_s/\Gamma_s = -0.14$ and 
$\Delta M_s=14.5\,{\rm ps}^{-1}$.}\label{fig3}
\end{figure}

\begin{figure}
$$\epsfxsize=0.5\textwidth\epsffile{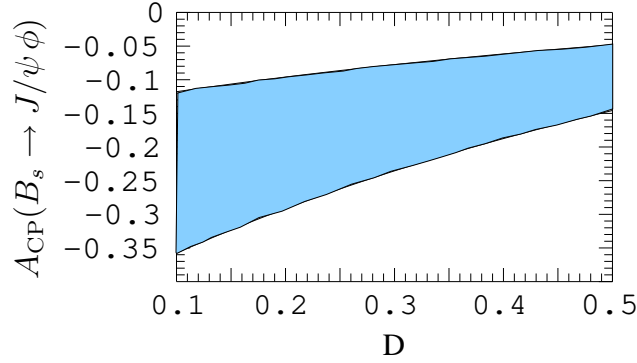}$$
\caption[]{The prediction of the SB--LR model for the CP-violating observable
$A_{\rm CP}(B_s\to J/\psi\,\phi)$ introduced in (\ref{ACP-def}) as a function
of the hadronic parameter $D$.}\label{fig4}
\end{figure}

Let us finally illustrate the CP-violating asymmetry (\ref{CP-asym}) 
of the decay $B_s\to J/\psi\,\phi$. In Fig.~\ref{fig3}, we plot this
CP asymmetry as a function of $t$, for fixed values of $D=0.3$, 
$\sin\phi_s = -0.38$, $\Delta \Gamma_s/\Gamma_s = -0.14$ and 
$\Delta M_s=14.5\,{\rm ps}^{-1}$. Although the $B^0_s$--$\overline{B^0_s}$ 
oscillations are very rapid, as can be seen in this figure, it should be 
possible to resolve them experimentally, for example at the LHC. The first 
extremal value of (\ref{CP-asym}), corresponding to $\Delta M_st =\pi/2$, 
is given to a very good approximation by
\begin{equation}\label{ACP-def}
A_{\rm CP}(B_s\to J/\psi\,\phi)=\left(\frac{1-D}{1+D}\right)\sin\phi_s,
\end{equation}
which would also fix the magnitude of the $B_s\to J/\psi\,\phi$ CP
asymmetry (\ref{CP-asym}) in the case of a negligible width difference 
$\Delta\Gamma_s$. In Fig.~\ref{fig4}, we show the prediction of the 
SB--LR model for (\ref{ACP-def}) as a function of the hadronic parameter 
$D$. For a value of $D=0.3$, the CP asymmetry may be as large as 
$-25\%$. The dilution through the hadronic parameter $D$ is not effective
in the case of the CP-violating observables of the $B_s\to J/\psi[\to l^+l^-]
\,\phi[\to K^+K^-]$ angular distribution, which allow us to probe
$\sin\phi_s$ directly \cite{DFN}.

\section{Conclusions}\label{sec:concl}
We have performed an analysis of mixing-induced CP-violating effects in 
$B_s\to D_s^+D^-_s$, $J/\psi\, \eta^{(')}$, $J/\psi\, \phi$ decays in the
SB--LR model with spontaneous CP violation, 
taking into account all presently available experimental 
constraints on the parameters of this model, and have demonstrated that 
the corresponding CP asymmetries may be as large as ${\cal O}(40\%)$, 
whereas the Standard Model predicts vanishingly small values. Since the 
decay amplitudes of these modes are not significantly affected in the 
SB--LR model, direct CP violation remains negligible, as in the Standard 
Model. From an experimental point of view, $B_s\to J/\psi\,\phi$ is a 
particularly promising mode, which is very accessible at $B$-physics 
experiments at hadron machines. We have proposed a simple strategy to 
search for indications of new physics in this transition, which does 
not require an angular analysis of the $J/\psi[\to l^+l^-]$ and 
$\phi[\to K^+K^-]$ decay products. In contrast to the large mixing-induced 
CP asymmetries in the $B_s$ channels, the SB--LR model predicts a small 
value for ${\cal A}^{\rm mix}_{\rm CP}(B_d\to J/\psi\, K_{\rm S})$ below 
$10\%$. Since the $B_s$ decays cannot be explored at the asymmetric 
$e^+$--$e^-$ $B$-factories operating at the $\Upsilon(4S)$ resonance, 
such a pattern would be in favour of hadronic $B$ experiments. We look 
forward to experimental data to check whether this scenario is actually 
realized by Nature.

\section*{Acknowledgements}

P.B.\ is supported by DFG through a Heisenberg fellowship.

\end{document}